%% file: main.tex
\documentclass[10pt,twocolumn,letterpaper]{article}

\usepackage[pagenumbers]{iccv} 

\def\name{Safe-VAR}

\usepackage{xcolor}

\input{preamble}

\usepackage{amsmath}
\usepackage{multicol}

\usepackage{textcomp}

\usepackage{amsthm}
\usepackage{amssymb}
\usepackage{amsmath}
\usepackage{arydshln} 

\usepackage{multirow}
\usepackage{booktabs}

\usepackage{algorithm}
\usepackage{algorithmic}
\definecolor{iccvblue}{rgb}{0.21,0.49,0.74}
\usepackage[pagebackref,breaklinks,colorlinks,allcolors=iccvblue]{hyperref}

\def\paperID{6325} 
\def\confName{ICCV}
\def\confYear{2025}

\title{Safe-VAR: Safe Visual Autoregressive Model for Text-to-Image Generative Watermarking}

\author{
    Ziyi Wang\textsuperscript{1}, 
    Songbai Tan\textsuperscript{2}, 
    Gang Xu\textsuperscript{6}, 
    Xuerui Qiu\textsuperscript{4}, 
    Hongbin Xu\textsuperscript{3}\\
    Xin Meng\textsuperscript{5}, 
    Ming Li\textsuperscript{6}, 
    Fei Richard Yu\textsuperscript{6} 
    \\
    \\
    \textsuperscript{1}Zhejiang University
    \textsuperscript{2}Shenzhen University 
    \textsuperscript{3}South China University of Technology\\
    \textsuperscript{4}Institute of Automation, Chinese Academy of Sciences
     \quad
    \textsuperscript{5}Peking University \\
    \textsuperscript{6}Guangdong Laboratory of Artificial Intelligence and Digital Economy (SZ)
}

\begin{document}
\maketitle
\input{sec/0_abstract}

\input{sec/1_intro}
\input{sec/2_related}
\input{sec/3_method}

\input{sec/4_experi}

\input{sec/5_conc}

\newpage 
{
    \small
    \bibliographystyle{ieeenat_fullname}
    \bibliography{main}
}

\input{sec/X_suppl}

\end{document}

%% file: preamble.tex
\usepackage{colortbl}

%% file: sec/0_abstract.tex
\begin{abstract}
With the success of autoregressive learning in large language models, it has become a dominant approach for text-to-image generation, offering high efficiency and visual quality. However, invisible watermarking for visual autoregressive (VAR) models remains underexplored, despite its importance in misuse prevention. Existing watermarking methods, designed for diffusion models, often struggle to adapt to the sequential nature of VAR models.
To bridge this gap, we propose \name{}, the first watermarking framework specifically designed for autoregressive text-to-image generation. Our study reveals that the timing of watermark injection significantly impacts generation quality, and watermarks of different complexities exhibit varying optimal injection times. 
Motivated by this observation, we propose an Adaptive Scale Interaction Module, which dynamically determines the optimal watermark embedding strategy based on the watermark information and the visual characteristics of the generated image. This ensures watermark robustness while minimizing its impact on image quality. Furthermore, we introduce a Cross-Scale Fusion mechanism, which integrates mixture of both heads and experts to effectively fuse multi-resolution features and handle complex interactions between image content and watermark patterns. 
Experimental results demonstrate that \name{} achieves state-of-the-art performance, significantly surpassing existing counterparts regarding image quality, watermarking fidelity, and robustness against perturbations.
Moreover, our method exhibits strong generalization to an out-of-domain watermark dataset QR Codes.
\end{abstract}

%% file: sec/1_intro.tex


\begin{figure}[!t]
    \centering
    \includegraphics[width=0.99\linewidth]{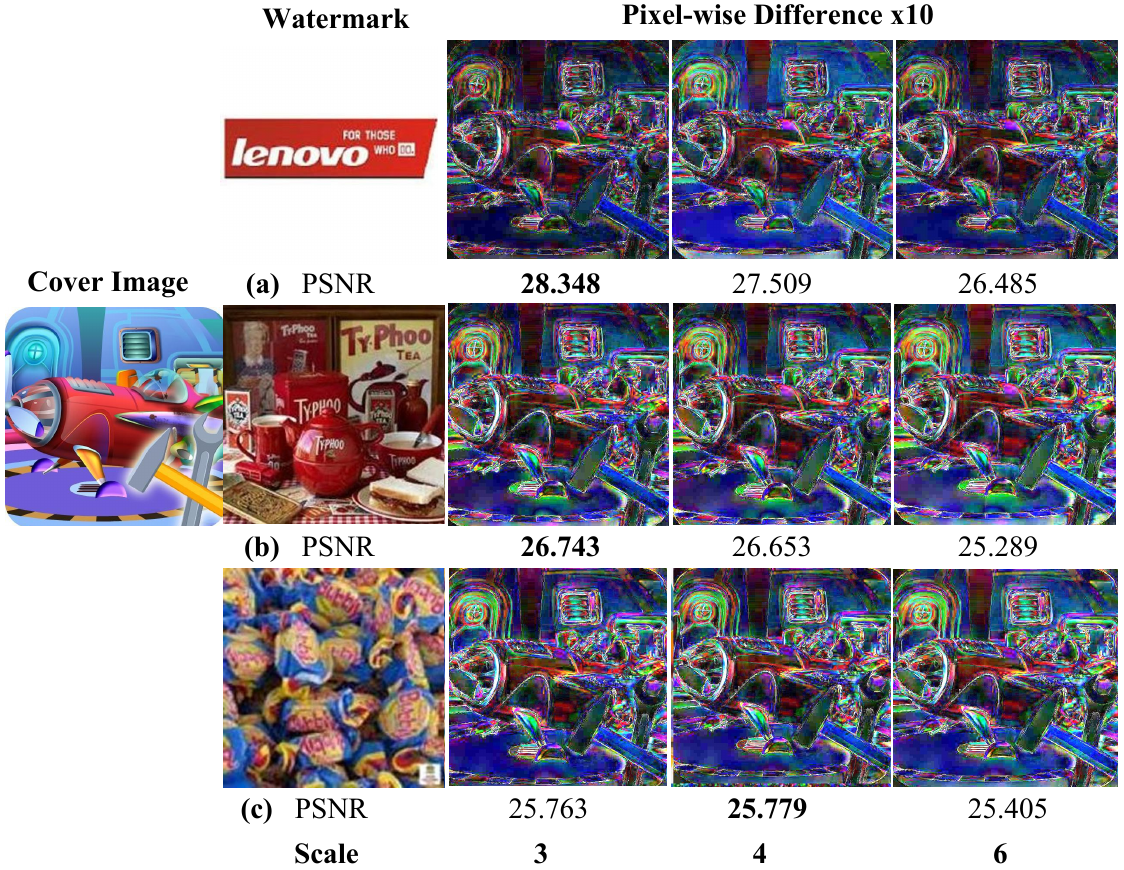}
    \vspace{-3mm}
    \caption{
    Watermarking efficacy of the same cover image with watermarks of varying complexity. (Pixel-wise Difference$\times 10$: Computed between the cover image and watermarked image, with differences scaled by 10 for enhanced visualization.) Manually selecting embedding scales leads to significant variability: the same watermark behaves differently across scales, while the same scale yields inconsistent results for different complexities. This highlights the need for adaptive scale selection in watermark embedding.
    }
    \label{fig: motivation}
    \vspace{-4mm}
\end{figure}

\section{Introduction}
\label{sec:intro}
Large language models (LLMs) have made significant strides in natural language processing. At their core, autoregressive (AR) models drive this progress with strong generalization and versatility \cite{tian2024visual}. These models generate text sequentially, predicting each token based on prior context. Notable examples include DeepSeek \cite{bi2024deepseek,liu2024deepseek}, the GPT series \cite{radford2018improving,radford2019language,ouyang2022training,achiam2023gpt}, and other leading LLMs \cite{chowdhery2023palm,anil2023palm,hoffmann2022training,touvron2023llama,le2023bloom,sun2021ernie,bai2023qwen,team2023gemini}, all of which have effectively leveraged this paradigm to achieve state-of-the-art performance.

Originally designed for NLP tasks, autoregressive (AR) models have recently shown strong potential in visual generation, including text-to-image, text-to-video, and image editing. AR-based visual models, such as LlamaGen \cite{sun2024autoregressive}, VAR \cite{tian2024visual}, and Hart \cite{tang2024hart}, employ visual tokenizers that use vector quantization (VQ) to map images from continuous pixel space to discrete visual tokens. These tokens are then processed using the same AR generation strategy as LLMs to synthesize images.


Compared to conventional diffusion models, AR visual generation models offer significantly higher computational efficiency and scalability. Diffusion models, while capable of producing high-quality images, require hundreds to a thousand iterative denoising steps, leading to substantial computational overhead. In contrast, AR models generate images by sequentially predicting tokens in much less steps, dramatically improving generation speed while maintaining competitive visual quality. Moreover, AR models can leverage the extensive pre-trained knowledge of LLMs \cite{radford2018improving, bi2024deepseek}, which operate under the same paradigm, further reducing the training cost and enhancing their generative capabilities.

Previous research on watermark protection falls into two main categories. The first focuses on post-processing, embedding watermarks after image generation \cite{hsu1999hidden,baluja2017hiding,xia1998wavelet,fei2022supervised,yu2021artificial,cui2023diffusionshield}, which compromises security by exposing the original image. The second integrates watermarks during generation in diffusion models, enhancing security but at a high computational cost \cite{zhao2023recipe,wen2023tree,meng2025latent,lei2024diffusetrace}. Some studies \cite{xiong2023flexible,ma2024safe, fernandez2023stable,kim2024wouaf} mitigate this by fine-tuning the visual decoder for direct watermark embedding, improving efficiency. However, due to fundamental differences in generation paradigms—diffusion models generate images in parallel, whereas AR models follow a sequential process—existing watermarking techniques cannot be directly applied to AR models. This underscores the urgent need for watermark protection algorithms tailored to AR models.

Recently, next-scale prediction-based AR models have demonstrated significant advantages in image generation \cite{kang2023scaling,tang2024hart}. In sequential multi-scale generation, AR models process tokens at different scales, each representing varying levels of image detail. We observe that embedding watermarks at different scales impacts image quality differently. As shown in \cref{fig: motivation}, embedding the same watermark at different scales results in varying degrees of distortion. Moreover, watermark complexity also influences distortion, even when embedded at the same scale, and the optimal embedding scale varies for different watermarks.

Based on these observations, we propose \name{}, the first watermarking framework specifically designed for text-to-image AR models. To achieve effective watermark embedding while maintaining image quality, \name{} introduces three key components. First, the Adaptive Scale Interaction Module dynamically selects the optimal watermark embedding strategy based on the complexity of both the watermark and the generated image. This ensures robust watermark embedding while minimizing visual degradation. Second, the Cross-Scale Fusion Module facilitates the integration of multi-scale watermark and image features by incorporating mixture of heads (MoH) \cite{jin2024moh} and mixture of experts (MoE) \cite{NEURIPS2021_48237d9f} structures. This design enables the model to efficiently process and optimize multimodal data, enhancing generation performance. Finally, the Fusion Attention Enhancement Module employ both spatial and channel attention mechanisms to effectively integrate image and watermark features across different scales. 
Extensive experiments on LAION-Aesthetics \cite{schuhmann2022laion}, LSUN-Church \cite{yu2015lsun}, and ImageNet \cite{deng2009imagenet} demonstrate that \name{} achieves state-of-the-art performance, significantly surpassing previous generative counterparts in image quality, watermark invisibility, and robustness against perturbations.

Our main contributions can be summarized as follows:
\begin{itemize}  
    \item We propose \name{}, the first watermarking protection algorithm specifically designed for AR visual generation models. It demonstrates significant superiority over existing text-to-image methods in terms of image quality, watermark invisibility, and robustness against perturbations. Additionally, we showcase the generalization ability of \name{} through zero-shot evaluation.  
    
    \item We introduce the Adaptive Scale Interaction Module, which dynamically selects the optimal watermark embedding strategy based on watermark and image complexity, ensuring robust embedding while minimizing visual distortion.  
    
    \item We design the Cross-Scale Fusion Module, which integrates mixture-of-heads (MoH) and mixture-of-experts (MoE) to effectively fuse multi-scale watermark and image features, enhancing embedding efficiency and stability.    
\end{itemize}


%% file: sec/2_related.tex
\begin{figure*}[t]
    \centering
     \includegraphics[width=0.99\linewidth]{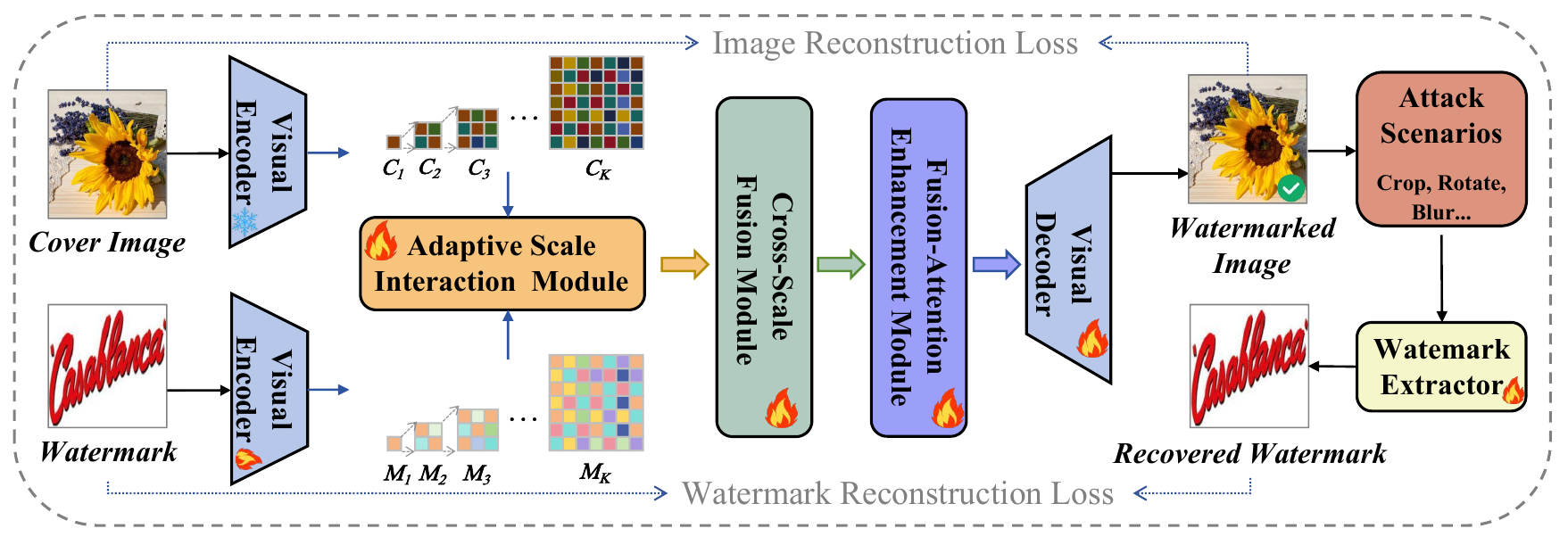}
    \caption{The overall pipeline of our Safe-VAR. Safe-VAR selects and fuses multi-scale residual maps for the watermark and image through adaptive scale interaction, cross-scale fusion, and fusion-attention enhancement, generating a refined feature representation, which is then decoded into the watermarked image. The Watermark Extractor enables reliable retrieval even under attack scenarios, optimizing both image quality and watermarking fidelity.}\label{fig: pipeline}
\end{figure*}
\section{Related Works}
\label{sec:formatting}
\paragraph{Text-to-Image Generation.}
The evolution of text-to-image generation has progressed through overcoming key technological limitations. Early GANs \cite{sauer2023stylegan, kang2023scaling} suffered from instability and poor controllability, while diffusion models \cite{baldridge2024imagen, ramesh2022hierarchical, rombach2022high, ho2022classifier, rombach2022high,podell2023sdxl}improved quality but faced high computational costs and inefficiency. Recently, autoregressive models \cite{tian2024visual, sun2024autoregressive} have emerged as a promising solution, leveraging LLM priors and sequential modeling to achieve an optimal balance between efficiency and quality. 

Autoregressive text-to-image generation methods, such as Parti \cite{yu2022scaling} and Emu3 \cite{wang2024emu3}, encode text and images as token sequences and employ a generative Transformer to predict image tokens from text tokens via next-token prediction.
Inspired by the global structure of visual information, Visual AutoRegressive modeling (VAR)\cite{tian2024visual} reformulates autoregressive modeling for image generation as a next-scale prediction framework, substantially enhancing generation quality and sampling efficiency. HART \cite{tang2024hart}, built upon VAR, introduces a hybrid tokenizer that incorporates both discrete and continuous tokens to enhance efficiency. Fluid \cite{fan2024fluid}proposes random-order models and replaces traditional discrete tokenizers with continuous tokenizers to further optimize performance. 
Their hierarchical decoding mechanism enhances image coherence, positioning them as a key technology for the next generation of text-to-image generation.

\paragraph{Image Watermarking.}

Watermarking techniques for AI-generated content can be broadly classified into post-hoc watermarking and generation-phase watermarking, based on the stage at which the watermark is embedded. 

Post-hoc watermarking methods \cite{xia1998wavelet,hsu1999hidden} introduce watermarks after image generation, typically through binary string embedding \cite{fei2022supervised,yu2021artificial,hsu1999hidden}, text-based encoding \cite{cui2023diffusionshield}, or graphical watermark injection \cite{ong2021protecting}. While these methods are applicable to generated images, they require additional processing steps that may degrade perceptual quality \cite{fernandez2023stable} and fail to prevent original image leakage.

Generation-phase watermarking integrates watermarks directly into the generation process, leveraging diffusion models to embed diffusion-native watermarks \cite{zhao2023recipe,fernandez2023stable,min2024watermark,xiong2023flexible,lei2024diffusetrace,meng2025latent,yang2024gaussian,ci2024ringid}. Tree-Ring \cite{wen2023tree} embeds watermarks in the initial noise, improving robustness but lacking multi-key identification capabilities \cite{ci2024ringid}. Alternative approaches, such as FSW \cite{xiong2023flexible}, StableSignature \cite{fernandez2023stable}, and WOUAF \cite{kim2024wouaf}, modify the Variational Autoencoder (VAE) decoder in Latent Diffusion Models (LDMs) to act as a watermark embedding module. However, binary watermarking remains vulnerable to DDIM inversion \cite{song2020denoising}, overlay attacks \cite{wang2019attacks}, and backdoor attacks \cite{chen2017targeted,gu2019badnets}. To enhance robustness, Safe-SD \cite{ma2024safe} encodes graphical watermarks within imperceptible structure-related pixels, ensuring resilience by binding watermark pixels across diffusion steps.


Despite the improved robustness and stealthiness of diffusion-native watermarking, its applicability remains constrained by model architecture, limiting adaptation to autoregressive models. 



%% file: sec/3_method.tex
\section{Methods}
\label{sec: method}


In this section, we first introduce the text-to-image Visual Autoregressive (VAR) model in Section~\ref{sec: Preliminaries}. 
We then present the key components of our \name{} in Section~\ref{sec: AISM}, \ref{sec: CSFM} and \ref{sec: FAEM}, followed by a detailed explanation of our training setup in Section~\ref{sec: Train}.

\subsection{Preliminaries}
\label{sec: Preliminaries}


Here, we briefly describe two main components of the visual autoregressive (VAR) model in image generation. 

\noindent\textbf{Multi-scale Vector Quantized Variational Autoencoder.}
The Multi-scale Vector Quantized Variational Autoencoder (VQVAE) is proposed by Tian~\textit{et al.}~\cite{tian2024visual} to encode an image to multi-scale discrete token maps, which are necessary for VAR learning.
In the encoding stage, it first encode an image $I$ into a feature map $F \in \mathbb{R}^{h \times w \times d}$. 
Then the feature map $F$ is quantized into $K$ multi-scale residual maps $\{R_1, R_2, \dots, R_K\}$, with each $R_k$ having a resolution of $h_k \times w_k$. 
In the reconstruction stage, the multi-scale residual maps are utilized to iteratively reconstruct an approximation of the original feature map $F$, as formulated in Equation~\eqref{eq:feature_approx}:
%
\begin{equation}
F_k = \sum_{i=1}^k \phi(R_i, (h, w)),
\label{eq:feature_approx}
\end{equation}
%
where $\phi(\cdot)$ is bilinear upsampling and $F_k$ denotes the cumulative sum of the upsampled residuals $\{R_1, R_2, \dots, R_k\}$.
Finally, we can obtain the reconstructed image by decoding $F_K$ with the decoder.

\noindent\textbf{Transformer.}
Within the VAR framework, the Transformer is designed to autoregressively predict the residuals \(R_k\) for the next scale, leveraging previous predictions $(R_1, \dots, R_{k-1})$ and the text input.
%
Formally, the autoregressive likelihood is defined as:
\begin{equation}
p(R_1, \dots, R_K) = \prod_{k=1}^K p(R_k \mid R_1, \dots, R_{k-1}, \boldsymbol{\Psi}(t)),
\label{eq:autoregressive_likelihood}
\end{equation}
where $\boldsymbol{\Psi}(t)$ is the text embedding from the text encoder. 
%

\begin{figure}[h]
    \centering
    \includegraphics[width=1.0\linewidth]{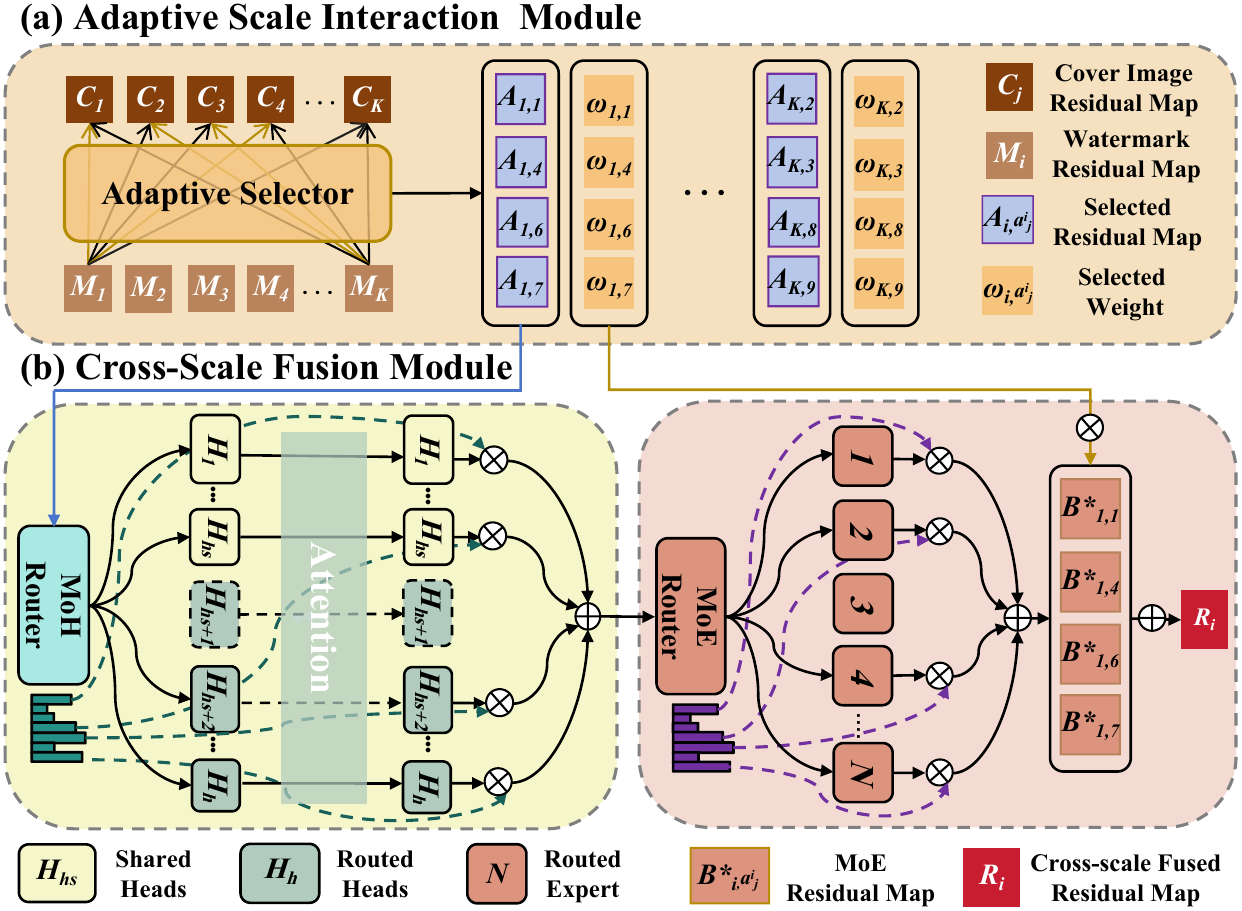}
    \caption{Structures of Adaptive Scale Interaction Module(a) and Cross-Scale Fusion Module(b). The Adaptive Selector in (a) dynamically selects the top-$k$ scales from the multi-scale residual maps of the cover image $\mathbb{C}$ and watermark $\mathbb{M}$, while also computing the corresponding weights. Then, in (b), the selected residual maps $A_{i,a^i_j}$ undergo cross-scale fusion of the watermark and image through the MoH and MoE routing mechanisms. The weights computed in (a) are then applied to the MoE residual maps $B^*_{i,a^i_j} $, ultimately generating the fused residual map $R_i$ for each scale $i$.}\label{fig: moudule}
\end{figure}

\subsection{Adaptive Scale Interaction Module}
\label{sec: AISM}


To achieve effective multi-scale integration of image and watermark residual maps while preserving image quality, we propose the Adaptive Scale Interaction Module (ASIM) which dynamically identifies the optimal combination of image and watermark residual maps based on their correlations.

As shown in \cref{fig: pipeline}, Leveraging the shared codebook $Z$, we first use the pre-trained Multi-scale VQVAE to encode both the cover image $I$ and the watermark image $W$, producing $K$ multi-scale residual maps for the cover Image, $\mathbb{C} = \{C_1, C_2, \dots, C_K\}$, and for the watermark image, $\mathbb{M} =\{M_1, M_2, \dots, M_K\}$.
Subsequently, each watermark residual map $M_i$ is concatenated with every cover image residual map $C_j$ to form $A_{i,j}$. 
This concatenated residual map is then pooled and processed by a multilayer perceptron (MLP) and Softmax operation to predict its selection weight $w_{i,j}$, as follows:
%
\begin{equation}
w_{i,j} = \text{Softmax}(\text{MLP}(A_{i,j})), \quad i, j \in \{1, 2, \dots, K\}.
\end{equation}
As illustrated in \cref{fig: moudule}, For each index $i \in \{1, 2, \dots, K\}$, we extract the subset of index $j \in \{1, 2, \dots, K\}$ according to the top-$k$ largest values of $w_{i,j}$, forming a set of index $\{a^i_1, a^i_2, \dots, a^i_k\}$.
%
%
Based on the $\{a^i_1, a^i_2, \dots, a^i_k\}$, each watermark residual map $M_i$ is concatenated with its top-$k$ relevant image residual maps $\{ C_{a^i_1}, C_{a^i_2},\dots C_{a^i_k} \}$, resulting in the set of selected residual maps $\{A_{i,a^i_1}, A_{i,a^i_2},\dots, A_{i,a^i_k}\}$, which serves as the foundation for the subsequent stage of information fusion.

\subsection{Cross-Scale Fusion Module}
\label{sec: CSFM}

Having obtained the optimal combination of image and watermark residual maps, we now focus on integrating the watermark and image features across multiple scales. 
To this end, we propose a Cross-Scale Fusion Module (CSFM) that employs both mixture of heads (MoH) and mixture of experts (MoE) mechanisms to realize effective multi-scale feature fusion, which is shown in \cref{fig: moudule}.

\paragraph{Mixture of Heads.}
Since the MoH leverages multi-head attention to capture intricate relationships among multi-scale features, it can enhancing the model's perception of both image and watermark information.
%
It is composed of $h$ attention heads, denoted by $\mathbb{H} = \{H_1, H_2, \dots, H_h \}$ along with a MoH router.
Within this structure, some attention heads are designated as shared heads that remain constantly active to extract common features across different scales, while routing heads are dynamically activated as needed to suit specific inputs.
%
The selected residual maps $A_{i,a^i_j}$, where $a^i_j$ belongs to the top-$k$ indices, is processed by the MoH router to select the most relevant attention heads.
Finally, MoH computes the query ($Q$), key ($K$), and value ($V$) matrices for each attention head and performs scaled dot-product attention to obtain the weighted sum to achieve MoH residual map:
\begin{equation}
B_{i,a^i_j}=\sum_{m=1}^h g_m H_m(W_o^m)=\mathrm{MoH}(A_{i,a^i_j}, A_{i,a^i_j}) ,
\label{eq:MoH_output}
\end{equation}
where $g_m$ denotes the routing score for each attention head, and $W_o^m$ is the projection matrix for transforming queries, keys, and values.

\paragraph{Mixture of Experts.}
The MoH residual maps $\{B_{i,a^i_1}, B_{i,a^i_2}, \dots, B_{i,a^i_k}\}$ are subsequently processed through several parallel MoE feedforward layers.
In conventional MoE architectures, the Transformer’s standard feedforward network (FFN) is replaced with MoE layers composed of multiple experts.
Typically, each token is assigned to one \cite{fedus2022switch} or two experts\cite{lepikhin2020gshard}.
%
In our design, a MoE router dynamically selects the appropriate experts for processing watermark and image features at various scales, ensuring efficient utilization of cross-scale information.
Every MoE residual map is computed as:
\begin{equation}
B^*_{i,a^i_j} = \sum_{n=1}^N \left( g_n\mathrm{MoE}_n \left(B_{i,a^i_j} \right) \right) + B_{i,a^i_j},
\end{equation}
where $N$ is the total number of experts, $g_n$ is the gating value for the $n$-th expert, and $\text{MoE}_n(\cdot)$ represents the $n$-th expert.

Finally, $\{B^*_{i,a^i_1}, B^*_{i,a^i_2}, \dots, B^*_{i,a^i_k}\}$ are aggregated using the selected weights $\{{w_{i, a^i_1}, w_{i, a^i_2}, \dots, w_{i, a^i_k}}\}$ to yield  the cross-scale fused residual map $R_{i}$:
\begin{equation}
R_{i}= \sum_{s=1}^{k} w_{i,a^{i}_{s}} \cdot B_{i,a^{i}_{s}}^*
\label{eq:final_fused_features}
\end{equation}

\subsection{Fusion Attention Enhancement Module}
\label{sec: FAEM}

To extract key information from the cross-scale fused residual maps $\{R_1, R_2, \dots, R_K\}$ and derive the final feature representation $F$, we propose the Fusion Attention Enhancement Module (FAEM). FAEM processes $\{R_1, R_2, \dots, R_K\}$ through a two-stage attention mechanism:
Spatial Attention refines spatial dependencies within the residual maps to enhance the input, while Channel Attention reweights the spatially enhanced features to emphasize critical information."
After the refinement, the processed $\{R_1, R_2, \dots, R_K\}$ are aggregated to form the final fused feature representation $F$, which is then fed into the decoder to generate the watermarked image.

\subsection{Network Training}
\label{sec: Train}

We train the network using multiple optimization objectives, each paired with its own loss functions. In the following sections, we provide a detailed description of these objectives and their associated loss functions.

\paragraph{Image Quality and Watermarking Fidelity.}


We train a U-Net-based extractor network $T(\cdot)$, producing $\hat{W} = T(\hat{I})$~\cite{ronneberger2015u} to extract the watermark from the watermarked image. 
The extractor employs a standard U-Net architecture with attention layers integrated into the bottleneck, thereby enhancing feature extraction.

To ensure the watermarked image $\hat{I}$ and watermark $\hat{W}$ closely resemble the original image $I$ and watermark $W$, we define the reconstruction loss function as:
\begin{equation}
\mathcal{L}_{\text{rec}} = \lambda_{\text{im}} \cdot | I - \hat{I} |^2 + \lambda_{\text{wm}} \cdot | W - \hat{W} |^2,
\end{equation}
where both terms use the L2 norm. The hyperparameters $\lambda_{\text{im}}$ and $\lambda_{\text{wm}}$ control the relative importance of the image and watermark reconstruction, with default values set to 1.

\paragraph{Balanced MoE Expert Selection.}
To prevent overtraining of specific experts in the MoE structure and ensure balanced expert selection, we apply the balanced load distribution mechanism across experts, inspired by \cite{fedus2022switch}. And the load balancing loss function is given by:
\begin{equation}
\mathcal{L}_{\text{MoE}} = N\cdot \sum_{i=1}^{N} \text{load}_{i} \cdot P_i,
\end{equation}
where $\text{load}_i$ represents the token fraction assigned to each expert, and $P_i$ is the routing probability for the expert $i$.

\paragraph{Overall Learning Objectives.}

The total loss function combines the reconstruction, perceptual, adversarial, and MoE balancing losses:
\begin{equation}
\begin{split}
\mathcal{L}_{\text{total}} = & \lambda_{\text{rec}} \cdot \mathcal{L}_{\text{rec}} + \lambda_{\text{perc}} \cdot \mathcal{L}_{\text{perc}} + 
\lambda_{\text{adv}} \cdot \mathcal{L}_{\text{adv}}
\\ & +
\lambda_{\text{MoE}} \cdot \mathcal{L}_{\text{MoE}},
\end{split}
\end{equation}
where $\mathcal{L}_{\text{perc}}$ is the perceptual loss based on LPIPS \cite{zhang2018unreasonable}, and $\mathcal{L}_{\text{adv}}$ is the adversarial loss computed using a PatchGAN discriminator \cite{isola2017image}, which is trained simultaneously with the image decoder. The hyperparameters control the weights of different losses, with default values set to $\lambda_{\text{perc}} = 0.1$ for perceptual loss, $\lambda_{\text{adv}} = 0.1$ for adversarial loss, and $\lambda_{\text{MoE}} = 0.02$ for MoE balancing loss.


\begin{table*}[ht]
\centering
\scriptsize
\renewcommand{\arraystretch}{1.2} 
\setlength{\tabcolsep}{2pt} 
\begin{tabular}{ccccccccccccccccc}
\toprule
\multirow{3}{*}{Methods} & \multirow{3}{*}{Type}                                                           & \multicolumn{15}{c}{\textbf{Cover/Watermarked image pair}}                                                                                          \\ \cline{3-17} 
                         &                                                                                 & \multicolumn{5}{c}{LAION-Aesthetic}   & \multicolumn{5}{c}{LSUN-Church}            & \multicolumn{5}{c}{ImageNet}               \\ \cline{3-17} 
                         &                                                                                 & PSNR$\uparrow$  & MAE$\downarrow$  & RMSE$\downarrow$  & SSIM$\uparrow$ & LIPIPS$\downarrow$ & PSNR$\uparrow$  & MAE$\downarrow$   & RMSE$\downarrow$  & SSIM$\uparrow$ & LIPIPS$\downarrow$ & PSNR$\uparrow$  & MAE$\downarrow$   & RMSE$\downarrow$  & SSIM$\uparrow$ & LIPIPS$\downarrow$ \\ \hline
HiDDeN                   & \multirow{2}{*}{\begin{tabular}[c]{@{}c@{}}Post-hoc,\\ graphics\end{tabular}}   & 31.683 & 5.281  & 7.030  & 0.962 & 0.144   & 32.723 & 4.706  & 6.142  & 0.965 & 0.165   & 32.291 & 5.064  & 6.642  & 0.960 & 0.162   \\
Baluja et.al.                  &                                                                                 & 28.634 & 7.411  & 10.230 & 0.952 & 0.196   & 30.561 & 6.061  & 7.858  & 0.960 & 0.207   & 29.874 & 6.568  & 8.842  & 0.954 & 0.203   \\ \hdashline

Stable Signature & \begin{tabular}[c]{@{}c@{}}Generation,\\ strings\end{tabular} 
& 25.326 & 10.517 & 14.543 & 0.787 & 0.205 
& 26.540 & 9.114 & 12.227 & 0.853 & 0.192 
& 24.627 & 12.121 & 15.448 & 0.764 & 0.213 \\ 
\hdashline

Safe-SD (256) & \multirow{2}{*}{\begin{tabular}[c]{@{}c@{}}Generation,\\ graphics\end{tabular}}  
&  \cellcolor{gray!30} 22.269 & \cellcolor{gray!30} 13.640 & \cellcolor{gray!30} 21.046 & \cellcolor{gray!30} 0.663 &\cellcolor{gray!30}  0.224 
& \cellcolor{gray!30} 21.070 &\cellcolor{gray!30}  14.746 &\cellcolor{gray!30}  23.617 &\cellcolor{gray!30}  0.631 & \cellcolor{gray!30} 0.223 
& \cellcolor{gray!30} 23.074 & \cellcolor{gray!30} 12.642 & \cellcolor{gray!30} 19.583 & \cellcolor{gray!30} 0.660 &\cellcolor{gray!30}  0.230 \\ 

Safe-SD (512) &  
& 24.586 & 10.490 & 16.552 & 0.731 & 0.207 
& 25.428 & 9.490 & 14.210 & 0.767 & 0.255 
& 24.984 & 10.254 & 16.039 & 0.712 & 0.228 \\ 
\hdashline

Safe-VAR (256) & \multirow{3}{*}{\begin{tabular}[c]{@{}c@{}}Generation,\\ graphics\end{tabular}}  
&  \cellcolor{gray!30} 24.573 & \cellcolor{gray!30} 10.666 & \cellcolor{gray!30} 15.668 & \cellcolor{gray!30} 0.779 & \cellcolor{gray!30} 0.151 
& \cellcolor{gray!30} 25.563 & \cellcolor{gray!30} 9.095 & \cellcolor{gray!30} 14.205 &\cellcolor{gray!30}  0.772 &\cellcolor{gray!30}  0.217 
& \cellcolor{gray!30} 24.984 & \cellcolor{gray!30} 10.254 &\cellcolor{gray!30}  16.039 &\cellcolor{gray!30}  0.712 & \cellcolor{gray!30} 0.228 \\ 

Safe-VAR (512) &  
& 28.417 & 6.809  & 10.301 & 0.854 & 0.124 
& 29.173 & 5.974  & 9.361  & 0.861 & 0.151 
& 28.453 & 6.985  & 10.738 & 0.807 & 0.163 \\ 
Safe-VAR (1024) &  
&  \cellcolor{yellow!30}32.526 &\cellcolor{yellow!30} 4.454 &\cellcolor{yellow!30} 6.661 &\cellcolor{yellow!30} 0.918 &\cellcolor{yellow!30} 0.104 
&\cellcolor{yellow!30} 32.438 &\cellcolor{yellow!30} 4.527 &\cellcolor{yellow!30} 6.674 &\cellcolor{yellow!30} 0.912 &\cellcolor{yellow!30} 0.105 
& \cellcolor{yellow!30}32.496 & \cellcolor{yellow!30}4.486 & \cellcolor{yellow!30}6.672 &\cellcolor{yellow!30} 0.915 &\cellcolor{yellow!30} 0.104 \\ 
\hline
\end{tabular}

\begin{tabular}{ccccccccccccccccc}
\hline
\multirow{3}{*}{Methods} & \multirow{3}{*}{Type}                                                           & \multicolumn{15}{c}{\textbf{Watermark/Recovered watermark image pair}}                                                                                       \\ \cline{3-17} 
                         &                                                                                 & \multicolumn{5}{c}{LAION-Aesthetic}  & \multicolumn{5}{c}{LSUN-Church}           & \multicolumn{5}{c}{ImageNet}              \\ \cline{3-17} 
                         &                                                                                 & PSNR$\uparrow$  & MAE$\downarrow$  & RMSE$\downarrow$  & SSIM$\uparrow$ & LIPIPS$\downarrow$ & PSNR$\uparrow$  & MAE$\downarrow$  & RMSE$\downarrow$  & SSIM$\uparrow$ & LIPIPS$\downarrow$ & PSNR$\uparrow$  & MAE$\downarrow$  & RMSE$\downarrow$  & SSIM$\uparrow$ & LIPIPS$\downarrow$ \\ \hline
HiDDeN                   & \multirow{2}{*}{\begin{tabular}[c]{@{}c@{}}Post-hoc,\\ graphics\end{tabular}}   & 29.145 & 6.312 & 10.863 & 0.887 & 0.257   & 29.567 & 5.618 & 10.016 & 0.903 & 0.214   & 30.265 & 5.443 & 9.112  & 0.898 & 0.236   \\
Baluja et.al.                   &                                                                                 & 26.893 & 8.459 & 13.639 & 0.875 & 0.261   & 27.663 & 7.874 & 12.162 & 0.892 & 0.225   & 27.241 & 8.289 & 12.939 & 0.881 & 0.249   \\ \hdashline
Safe-SD(256)             & \multirow{2}{*}{\begin{tabular}[c]{@{}c@{}}Generation,\\ graphics\end{tabular}} & \cellcolor{gray!30}  24.302 & \cellcolor{gray!30} 9.375 & \cellcolor{gray!30} 17.053 & \cellcolor{gray!30} 0.844 & \cellcolor{gray!30} 0.145   &\cellcolor{gray!30}  24.324 &\cellcolor{gray!30}  9.167 & \cellcolor{gray!30} 16.923 &\cellcolor{gray!30}  0.853 &\cellcolor{gray!30}  0.144   &\cellcolor{gray!30}  24.299 &\cellcolor{gray!30}  9.455 &\cellcolor{gray!30}  17.042 &\cellcolor{gray!30}  0.841 &\cellcolor{gray!30}  0.146   \\
Safe-SD(512)             &                                                                                 & 28.815 & 6.052 & 10.138 & 0.909 & 0.130   & 28.313 & 6.286 & 10.611 & 0.913 & 0.133   & 28.976 & 6.028 & 9.967  & 0.906 & 0.130   \\ \hdashline
Safe-VAR(256)            & \multirow{3}{*}{\begin{tabular}[c]{@{}c@{}}Generation,\\ graphics\end{tabular}} &  \cellcolor{gray!30} 28.218 &\cellcolor{gray!30}  6.082 &\cellcolor{gray!30}  10.642 &\cellcolor{gray!30}  0.901 &\cellcolor{gray!30}  0.102   &\cellcolor{gray!30}  28.359 &\cellcolor{gray!30}  5.827 &\cellcolor{gray!30}  10.413 & \cellcolor{gray!30} 0.911 &\cellcolor{gray!30}  0.100   &\cellcolor{gray!30}  28.263 &\cellcolor{gray!30}  6.090 &\cellcolor{gray!30}  10.555 &\cellcolor{gray!30}  0.900 &\cellcolor{gray!30}  0.101   \\
Safe-VAR(512)            &                                                                                 & 33.954 & 3.360 & 5.378  & 0.943 & 0.086   & 33.917 & 3.215 & 5.385  & 0.949 & 0.086   & 33.914 & 3.398 & 5.393  & 0.941 & 0.086   \\
Safe-VAR(1024)           &                                                                                 &  \cellcolor{yellow!30}39.721   &  \cellcolor{yellow!30}1.850     & \cellcolor{yellow!30}  2.742     & \cellcolor{yellow!30} 0.980     &\cellcolor{yellow!30}   0.058      &\cellcolor{yellow!30}    39.604    &\cellcolor{yellow!30}  1.886     &\cellcolor{yellow!30}    2.788    &\cellcolor{yellow!30}  0.979     &\cellcolor{yellow!30}    0.060     & \cellcolor{yellow!30}  39.712     & \cellcolor{yellow!30} 1.862     &\cellcolor{yellow!30}     2.763   &\cellcolor{yellow!30}  0.980     &\cellcolor{yellow!30}  0.059      \\ \bottomrule
\end{tabular}

\caption{Quantitative comparison of image quality and watermarking fidelity on different datasets. The upper section compares image quality between cover and watermarked image pairs, while the lower section evaluates  watermarking fidelity.“↑”: the larger the better, “↓”: the smaller the better.}
\label{tab: evalu}
\end{table*}

%% file: sec/4_experi.tex
\begin{figure*}
    \centering
    \includegraphics[width=0.9\linewidth]{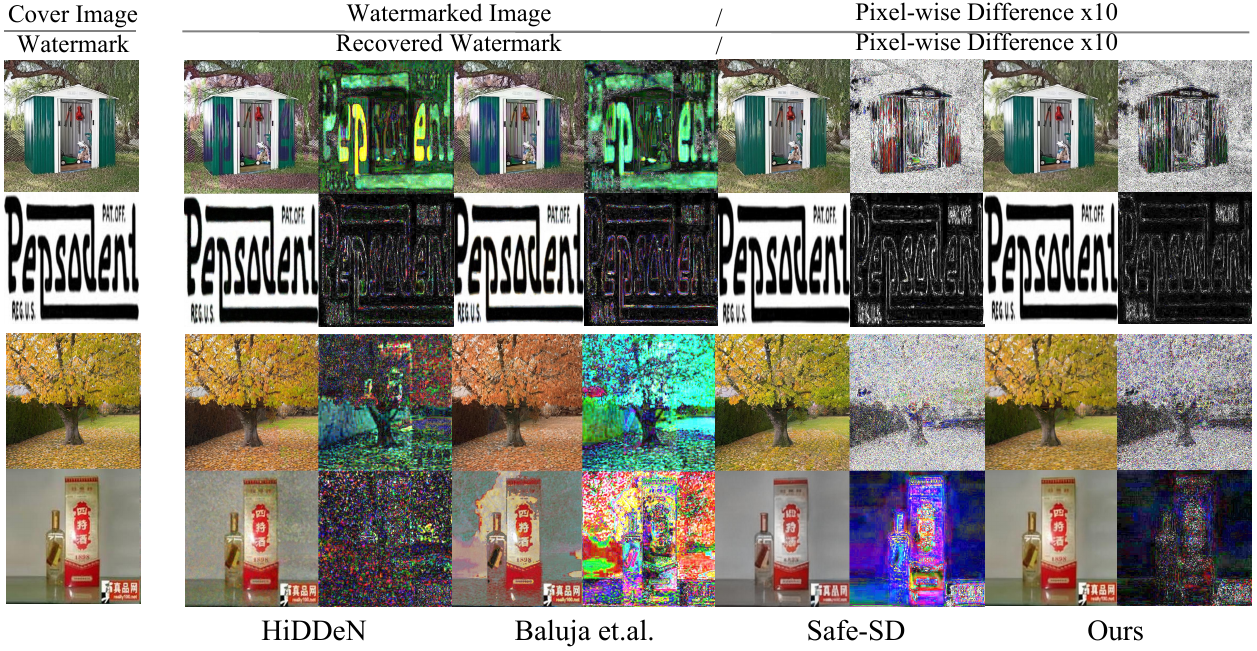}
    \caption{Qualitative comparison of Safe-VAR and baseline on LAION-Aesthetics (Pixel-wise differences$\times 10$: they are multiplied by a factor of 10 for better view). We can observe that our method maintains high image quality and watermarking fidelity .}
    \label{fig: comparison}
\end{figure*}

\section{Experiments}

In this part, we first compare \name{} with other watermarking protection models in Section \ref{visual quality}, presenting both quantitative and qualitative results. In Section \ref{Watermark Robustness} we evaluate the watermark robustness of Safe-VAR by testing its performance under different attack scenarios. In Section \ref{ablation studies}, we report a series of ablation experiments to verify the effectiveness of each component. The additional ablation studies and the experimental setting details are provided in \cref{ablation2} and \cref{experimental setting}.

\subsection{Image generation quality for watermarking}
\label{visual quality}

\begin{figure*}[th]
    \centering
    \includegraphics[width=0.9\linewidth]{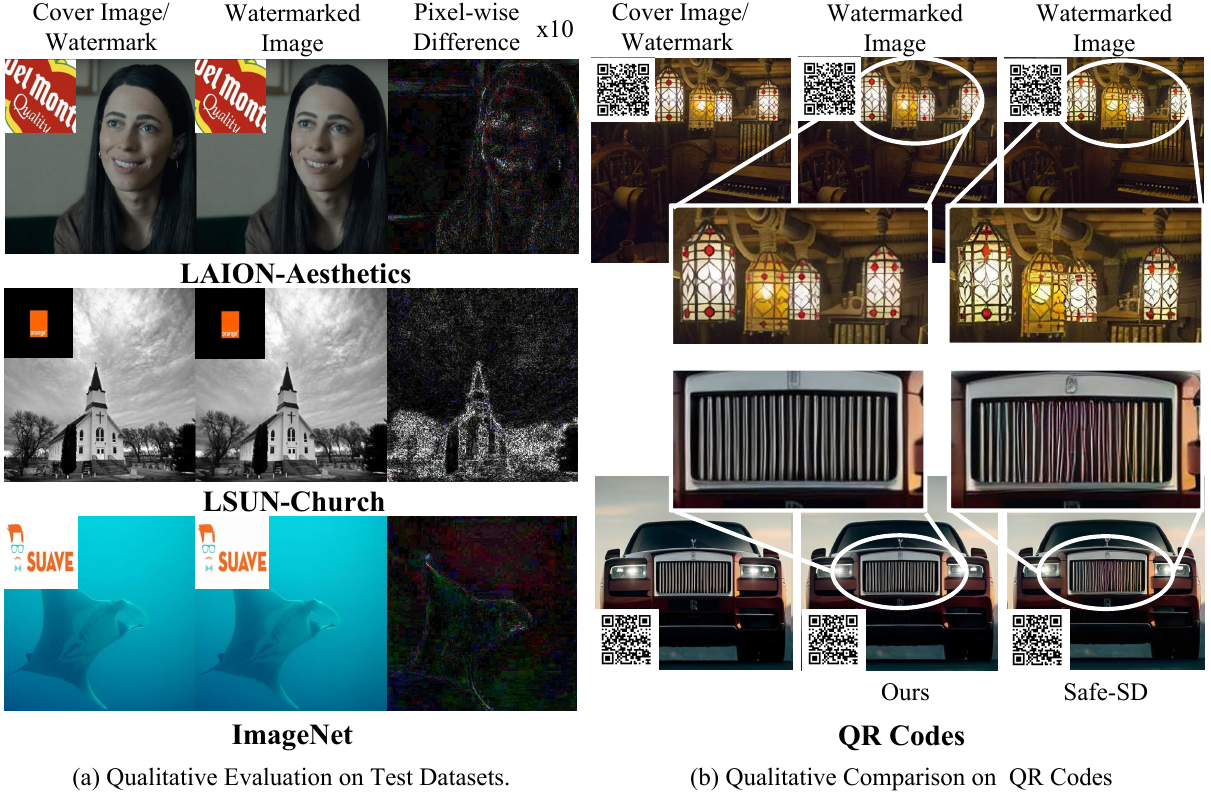} 
    \caption{Qualitative results on diverse datasets. (a) Evaluation on LAION-Aesthetics, LSUN-Church, and ImageNet, demonstrating the strong generalization ability of our method. (b) Zero-shot out-of-domain testing on a QR Codes dataset, showing that our method preserves more fine-grained details compared to Safe-SD, highlighting its robustness in unseen domains.}
    \label{fig:dataset}
\end{figure*}

\paragraph{Quantitative Evaluation.}
To evaluate the performance of our approach, we conduct a quantitative comparison between Safe-VAR and various watermarking protection methods, including HiDDeN \cite{hsu1999hidden}, Baluja et. al. \cite{baluja2017hiding}, Stable Signature \cite{fernandez2023stable}, and Safe-SD \cite{ma2024safe}. Among them, HiDDeN and Baluja et. al. adopt post-pocessing image watermarking protection methods, while Stable Signature and Safe-SD employ generative string watermarking and generative image watermarking techniques, respectively. Following \cite{li2024purified}, we train our model on LAION-Aesthetic \cite{schuhmann2022laion} and conduct evaluations on LAION-Aesthetic, LSUN-Church \cite{yu2015lsun}, and ImageNet \cite{deng2009imagenet}. For a more comprehensive evaluation, we selected several metrics including Peak Signal-to-Noise Ratio (PSNR), Mean Absolute Error (MAE), Root Mean Square Error (RMSE), Structural Similarity Index Measure (SSIM), and Learned Perceptual Image Patch Similarity (LIPIPS). All methods were tested on an image size of 512, and for Safe-sd and Safe-VAR, we also tested their performance on other image sizes. 

Table \ref{tab: evalu} shows the comparison results across three datasets. From this, we have the following findings. First, for watermarked images, our \name{} outperforms most methods on multiple datasets. Specially, \name{} outperforms generative watermarking methods, including Stable Signature and Safe-SD, across all five metrics in terms of the visual quality. Second, For recovered watermark images, our method surpasses all methods, including both post-embedding and generative watermarking methods. This indicates that our method not only ensures minimal degradation of the cover image quality but also achieves more accurate watermark extraction. Third, compared to other generative watermarking methods, our approach exhibits a more pronounced improvement in multiple resolutions. For the resolution of 256, compared to Safe-SD, Safe-VAR achieves 10.3\% and 17.5\% higher PSNR and SSIM, respectively, in cover/watermarked image pair tests on LAION-Aesthetic. For the resolution of 512, Safe-VAR outperforms Safe-SD in cover/watermarked image pair tests on LAION-Aesthetic, achieving 15.6\% and 16.8\% higher PSNR and SSIM, respectively.

\begin{table*}[ht]
    \centering
    \renewcommand{\arraystretch}{1.2} 
    \setlength{\tabcolsep}{5pt} 
    \resizebox{\linewidth}{!}{
    \begin{tabular}{lccccccccc}
        \toprule
        \cmidrule(lr){3-10}
        Method & Metrics & Crop 30\% & Rotate $\pm$ 40 & Blur 9 & Bright $\pm$ 0.4 & Noise 0.3 & Erase 30\% & Jpeg 50 & Combi. \\ 
        \midrule
        \multirow{3}{*}{Safe-SD} & FID $\downarrow$ & 5.842 & 6.4744 & 5.796 & 5.870 & 6.734 & 5.808 & 5.882 & 5.823 \\ 
        & PSNR $\uparrow$ & 28.097 & 28.370 & 28.263 & 28.133 & 26.234 & 27.907 & 28.146 & 27.859 \\ 
        & LPIPS $\downarrow$ & 0.134 & 0.134 & 0.135 & 0.135 & 0.147 & 0.133 & 0.137 & 0.136 \\ 
        \hdashline
        \multirow{3}{*}{Safe-VAR} & FID $\downarrow$ & 3.368 & 3.179 & 3.063 & 3.195 & 4.598 & 3.097 & 3.346 & 3.087 \\ 
        & PSNR $\uparrow$ & 32.686 & 32.865 & 33.037 & 32.861 & 32.620 & 33.086 & 32.388 & 33.612 \\ 
        & LPIPS $\downarrow$ & 0.089 & 0.090 & 0.088 & 0.090 & 0.094 & 0.086 & 0.093 & 0.084 \\ 
        \bottomrule
    \end{tabular}}
    \caption{Robustness comparison on LAION-Aesthetics dataset under various perturbations. Our method outperforms Safe-SD across all perturbations, demonstrating greater stability and stronger resistance to interference, especially under combined perturbations and noise-based distortions.}
    \label{tab: robustness}
\end{table*}

\paragraph{Qualitative Evaluation.}
To evaluate the image generation quality and watermarking fidelity, we conduct the qualitative comparions with other methods by visualizing the pixel-level differences between the cover image and watermarked image on LAION-Aesthetics, which are presented in \cref{fig: comparison}. Based on the results presented, we can obtain two conclusions. First, Safe-VAR preserves high visual quality in watermarked images, whereas post-processing methods like HiDDeN and Baluja et al. produce visible watermarks and significant color distortions. Our method mainly introduces subtle pixel-level differences$\times 10$, remaining imperceptible and traceability. Second, our method maintains the highest watermarking fidelity. Even for challenging images and complex watermarks (as shown in the second row), the recovered watermark finely preserves text details, whereas other methods exhibit significant loss and distortion. 

Moreover, we evaluate \name{} on LAION-Aesthetics (Top), LSUN-Church (Mid), and ImageNet (Bottom) to rigorously assess its generalization capability, as shown in \cref{fig:dataset}(a). The results demonstrate that our method consistently maintains superior visual quality across diverse datasets. Specifically, we perform zero-shot out-of-domain testing on a QR Codes dataset \cite{coledie_qr_codes} and compare it with Safe-SD. As illustrated in \cref{fig:dataset}(b), our method preserves significantly more fine-grained details of the original image compared to Safe-SD, highlighting its strong generalization ability and robustness in unseen domains. 

These findings underscore the effectiveness of our approach in achieving an optimal balance between image quality and watermarking fidelity while demonstrating exceptional adaptability across various datasets.

\subsection{Watermark Robustness}
\label{Watermark Robustness}

To rigorously evaluate the robustness of our graphical watermarking method, we follow WOUAF~\cite{kim2024wouaf} and apply eight different post-processing transformations to watermarked images. The transformations employed include cropping (30\%), random rotation ($\pm 40^\circ$), Gaussian blurring (kernel size = 9), brightness adjustment ($\pm 0.4$), additive Gaussian noise ($\sigma = 0.3$), random erasing (30\% region), JPEG compression (quality = 50), and composite perturbations (a random subset applied with 50\% probability).

From the results in Table \ref{tab: robustness}, Safe-VAR consistently outperforms Safe-SD across perturbation scenarios, exhibiting exceptional robustness, especially against composite attacks and noise-based distortions. Notably, Safe-SD exhibits significant performance degradation under noise perturbations, whereas our method achieves a PSNR of 32.620 in extracted watermarked images. These results substantiate the effectiveness and reliability of Safe-VAR in handling real-world distortions, reinforcing its superiority in robust graphical watermarking.

\subsection{Ablation Study}
\label{ablation studies}

As discussed in Section~\ref{sec: method}, the proposed \name{} consists of three key components: ASIM, CSFM, and FAEM. In this section, we conduct ablation studies to examine the impact of these modules on overall performance. 

\paragraph{Effect of Each component.}
Table \ref{tab: eachcom} presents the specific results of component ablation. The “Methods” column presents the learned variants of \name{} by removing one or three components of the standard \name{}. In all cases, the absence of a single module results in an obvious performance degradation. For example, the absence of the ASIM reduces image PSNR and SSIM by 2.9\% and 2.4\%. Similarly, the removal of CSFM decreases PSNR and SSIM by 3.8\% and 3.5\%. If all components are removed, the performance will be further degraded, as shown by the “Baseline”.  From the above results, we can conclude that three key components play a positive role in obtaining good performance and their cooperation further improves the overall performance.
\begin{table}[ht]
\centering
\renewcommand{\arraystretch}{1.3} 
\setlength{\tabcolsep}{3pt} 
\resizebox{0.9\linewidth}{!}{
\begin{tabular}{ccccc}
\toprule
Methods  & PSNR$\uparrow$       & MAE$\downarrow$      & RMSE$\downarrow$      & SSIM$\uparrow$     \\ \hline
Baseline & 27.02/26.77 & 8.04/7.84 & 12.44/9.98 & 0.80/0.88 \\
w/o ASIM & 27.59/32.45 & 7.43/4.02 & 11.15/6.94 & 0.83/0.91 \\
w/o CSFM & 27.35/32.25 & 7.64/4.13 & 11.23/6.12 & 0.82/0.90 \\
w/o FAEM & 28.31/33.38 & 6.93/3.67 & 10.74/5.42 & 0.84/0.92 \\
Safe-VAR & \textbf{28.42}/\textbf{33.95} & \textbf{6.81}/\textbf{3.36} & \textbf{10.30}/\textbf{5.38} & \textbf{0.85}/\textbf{0.94} \\ \bottomrule
\end{tabular}}
\caption{Ablation study on the impact of each component of the proposed \name{}. Each metric is presented as “image quality/watermarking fidelity.}
\label{tab: eachcom}
\end{table}

%% file: sec/5_conc.tex
\section{Conclusions}

In this paper, we propose \name{}, the first watermarking protection algorithm specifically designed for VAR generation models. Our method significantly outperforms existing text-to-image watermarking approaches in image quality, watermarking fidelity, and robustness. Furthermore, zero-shot evaluations demonstrate its strong generalization ability across diverse datasets.
To achieve imperceptible yet robust watermark embedding, we introduce the Adaptive Scale Interaction Module, which optimally selects residual map strategies based on image and watermark complexity. Additionally, the Cross-Scale Fusion Module leverages MoH and MoE to enhance multi-scale fusion stability. 
By exploring cutting-edge AR watermarking techniques, our method provides an innovative solution for copyright protection in AI-generated content.


%% file: sec/X_suppl.tex
\clearpage
\setcounter{page}{1}
\maketitlesupplementary

\appendix
\section{Experimental Setting}
\label{experimental setting}

\paragraph{Datasets.}

We use graphical logos from Logo-2K\cite{wang2020logo}  as predefined watermarks. For the image dataset, we train the model on 200,000 images from LAION-Aesthetics, a subset of LAION 5B \cite{schuhmann2022laion}, while LSUN-Church \cite{yu2015lsun} and ImageNet \cite{deng2009imagenet} are used as test sets to evaluate generalization. During testing, 5,000 images and 5,000 watermarks are randomly paired to perform the quantitatively experimental evaluations.

\paragraph{Implementation Details.}
Our training process follows a two-stage approach. In the first stage, we train the model at a $256 \times 256$ resolution for 40 epochs with a batch size of 60 and a learning rate of $1\times10^{-4}$, keeping the image encoder and codebook frozen while optimizing the remaining components. At 10,000 training steps, the discriminator is introduced, enabling it to refine the model’s ability to distinguish watermarked from cover images. In the second stage, we fine-tune the model at $512 \times 512$ and $1024 \times 1024$ resolutions for one epoch each, freezing the watermark visual encoder while continuing to optimize the other components. The batch size is reduced to 5, and the learning rate is adjusted to $5\times10^{-6}$. As in the first stage, the discriminator is activated at 10,000 steps, further enhancing its ability to improve watermark detection in high-resolution images.

\paragraph{Baseline.}
As no existing VAR-based generative watermarking framework, we compare our approach with several representative baseline methods, comprising two state-of-the-art (SOTA) post-hoc approaches and two diffusion-native watermarking techniques. For post-hoc watermarking, we evaluate HiDDeN \cite{hsu1999hidden}, the first deep learning-based method for embedding strings into images, which we modify to graphic watermark injection for consistency. We also include Baluja et al. \cite{baluja2017hiding}, which injects graphic watermarks, using the official implementation. For diffusion-native watermarking, we evaluate Stable Signature \cite{fernandez2023stable}, a representative text-to-image watermarking method, modifying its training process to randomly embed different strings into images. Additionally, we evaluate Safe-SD \cite{ma2024safe}, which implants the graphical watermarks into the imperceptible structure-related pixels. Due to the lack of publicly available training code and model weights, we reimplement it based on the descriptions of the original article.

\section{Additional Ablation Studies}
\label{ablation2}

\subsection{Effect of Adaptive Selector}
The observation in \cref{fig: motivation} underscores the crucial role of scale selection in determining the impact of watermarking on image quality. Bellow, we conduct an in-depth analysis from two perspectives to validate the effect of adaptive selector in the ASIM.

\begin{table}[ht]
\centering
\small
\renewcommand{\arraystretch}{1.3} 
\setlength{\tabcolsep}{3pt} 
\begin{tabular}{ccccc}
\toprule
Methods & PSNR$\uparrow$       & MAE$\downarrow$      & RMSE$\downarrow$      & SSIM$\uparrow$     \\ \hline
$K=2$  & 28.04/31.65 & 8.04/7.84 & 12.44/9.98 & 0.80/0.88 \\
$K=4$  & \textbf{28.42}/33.95 & 6.81/3.36 & \textbf{10.31}/\textbf{5.38} & 0.85/\textbf{0.94} \\
$K=8$  & \textbf{28.42}/\textbf{33.98} & \textbf{6.79}/\textbf{3.35} & \textbf{10.31}/5.39 & \textbf{0.86}/\textbf{0.94} \\
$K=14$ & 28.29/32.91 & 7.17/4.61 & 11.36/6.49 & 0.83/0.92 \\ \bottomrule
\end{tabular}
\caption{Performance metrics for different numbers of selected scales $K$. Each metric is presented as “image quality/watermarking fidelity.}
\label{tab: scale number}
\end{table}

\paragraph{The Number of Selected Scales.} Given 14 available scales, we analyze the impact of varying the number of selected scales $K$. As shown in Table \ref{tab: scale number}, increasing $K$ from 2 to 4 significantly improves PSNR, MAE, RMSE, and SSIM. However, further increasing $K$ to 8 does not yield additional benefits, as the performance remains nearly identical to that of $K = 4$. When $K$ is increased to 14, a slight performance degradation is observed, likely due to the introduction of redundancy or interference. Therefore, we determine $K=4$ as the optimal choice for scale fusion.

\begin{table}[ht]
\centering
\small
\renewcommand{\arraystretch}{1.3} 
\setlength{\tabcolsep}{3pt} 
\resizebox{\linewidth}{!}{
\begin{tabular}{ccccc}
\toprule
Methods   & PSNR$\uparrow$       & MAE$\downarrow$      & RMSE$\downarrow$       & SSIM$\uparrow$     \\ \hline
$k=0,3,6,9$ & 26.93/32.32 & 8.37/4.08 & 12.05/6.47  & 0.84/\textbf{0.94} \\
$k=3,6,9,12$ & 26.90/32.22 & 8.40/4.14 & 12.08/6.56  & 0.84/\textbf{0.94} \\
$k=0,1,2,3$ &27.18/32.37 & 8.06/4.07 & 11.74/6.44  & 0.83/0.93\\
$k=10,11,12,13$ &26.75/32.07 & 8.59/4.22 & 12.28/6.68  & 0.83/\textbf{0.94} \\
$k=\text{Top4}$    & \textbf{28.42}/\textbf{33.95} & \textbf{6.81}/\textbf{3.36} & \textbf{10.30}/\textbf{5.38}  & \textbf{0.85}/\textbf{0.94} \\ \bottomrule
\end{tabular}}
\caption{Comparison of different scale selection strategies. $k$ denotes the selected scale. Adaptive selection of the top four most effective scales ($k= \text{Top4}$) yields the best performance, highlighting the importance of adaptive scale selection.}
\label{tab: scale selection}
\end{table}

\paragraph{Fixed Scales Fusion \textit{vs} Adaptive Selection Scales Fusion.} Since we have determined that selecting four scales is optimal, the choice of which four scales significantly impacts the watermarking effect. Table \ref{tab: scale selection} demonstrate that the fixed scales fusion at equal intervals (e.g., $k = 0,3,6,9$ or $k = 3,6,9,12$) performs poorly, with lower PSNR and SSIM compared to the adaptive selection scales fusion. 
Similarly, selecting only the smallest scales ($k = 0,1,2,3$) does not bring significant improvements, while choosing the largest scales ($k = 10,11,12,13$) further degrades performance.
In contrast, adaptively selecting the four most effective scales ($k = \text{Top4}$) achieves the best performance across all metrics, further validating the importance of adaptive scale selection based on image complexity and watermark characteristics.

\subsection{Effect of Expert Numbers}
In watermark embedding, multi-scale token maps fusion and expert selection significantly impact generation quality. MoH facilitates feature interaction through shared heads and routed heads, where the ratio of shared heads may affect fusion capability. MoE improves model adaptability through dynamic expert selection, where the number of experts may influence generation performance and computational efficiency. To investigate these factors, we conducted the following two studies in different configurations.

\begin{table}[ht]
\centering
\small
\renewcommand{\arraystretch}{1.3} 
\setlength{\tabcolsep}{3pt} 
\begin{tabular}{ccccc}
\toprule
Methods        & PSNR$\uparrow$       & MAE$\downarrow$      & RMSE$\downarrow$      & SSIM$\uparrow$     \\ \hline
$h_s=2$, $h_r=4$ & 28.38/33.74 & 6.83/3.41 & 10.39/5.42 & 0.85/0.93 \\
$h_s=3$, $h_r=3$ & \textbf{28.42}/\textbf{33.95} & \textbf{6.81}/\textbf{3.36} & \textbf{10.30}/\textbf{5.38} & \textbf{0.85}/\textbf{0.94} \\ \hdashline
$N=1$         & 28.42/\textbf{33.95} & \textbf{6.81}/\textbf{3.36} & \textbf{10.30}/\textbf{5.38} & 0.85/\textbf{0.94} \\
$N=2$         & \textbf{28.43}/33.94 & \textbf{6.81}/3.38 & 10.32/5.43 & \textbf{0.86}/0.93 \\
$N=3$         & 28.36/33.87 & 6.89/3.54 & 10.35/5.49 & 0.83/0.92 \\
$N=4$         & 28.34/33.86 & 6.91/3.60 & 10.40/5.51 & 0.82/0.91 \\ \bottomrule
\end{tabular}
\caption{Ablation results of the MoH and MoE. The upper section examines the impact of the shared ($h_s$) and routed ($h_r$) head ratio in MoH, selecting 6 out of 8 heads for fusion. The lower section evaluates the effect of expert number ($N$) in MoE, selecting up to 4 experts. Each metric is reported as “image quality/watermarking fideity.”}
\label{tab: csfm}
\end{table}

\paragraph{Effect of Shared Head Ratio in MoH.} Following the MoH\cite{jin2024moh}, we use 8 attention heads, selecting 6 as shared and routed heads, to examine how different allocations affect watermark embedding performance.  As shown in \cref{tab: csfm}, setting $h_s = 3$, $h_r = 3$ achieves the best performance across PSNR, MAE, RMSE, and SSIM. 

\paragraph{Effect of the Number of Experts in MoE.}We evaluate the impact of the different selecting expert numbers from a total of 4 on generation quality and computational cost. \cref{tab: csfm} shows that $N = 1$ and $N = 2$ yield nearly identical results, while increasing $N$ to 3 or 4 slightly degrades performance. This indicates that adding more experts does not bring additional benefits and may introduce computational redundancy or overfitting. Thus, we set $N = 1$ to balance between performance and efficiency.

\begin{figure*}[ht]
    \centering
    \includegraphics[width=0.99\linewidth]{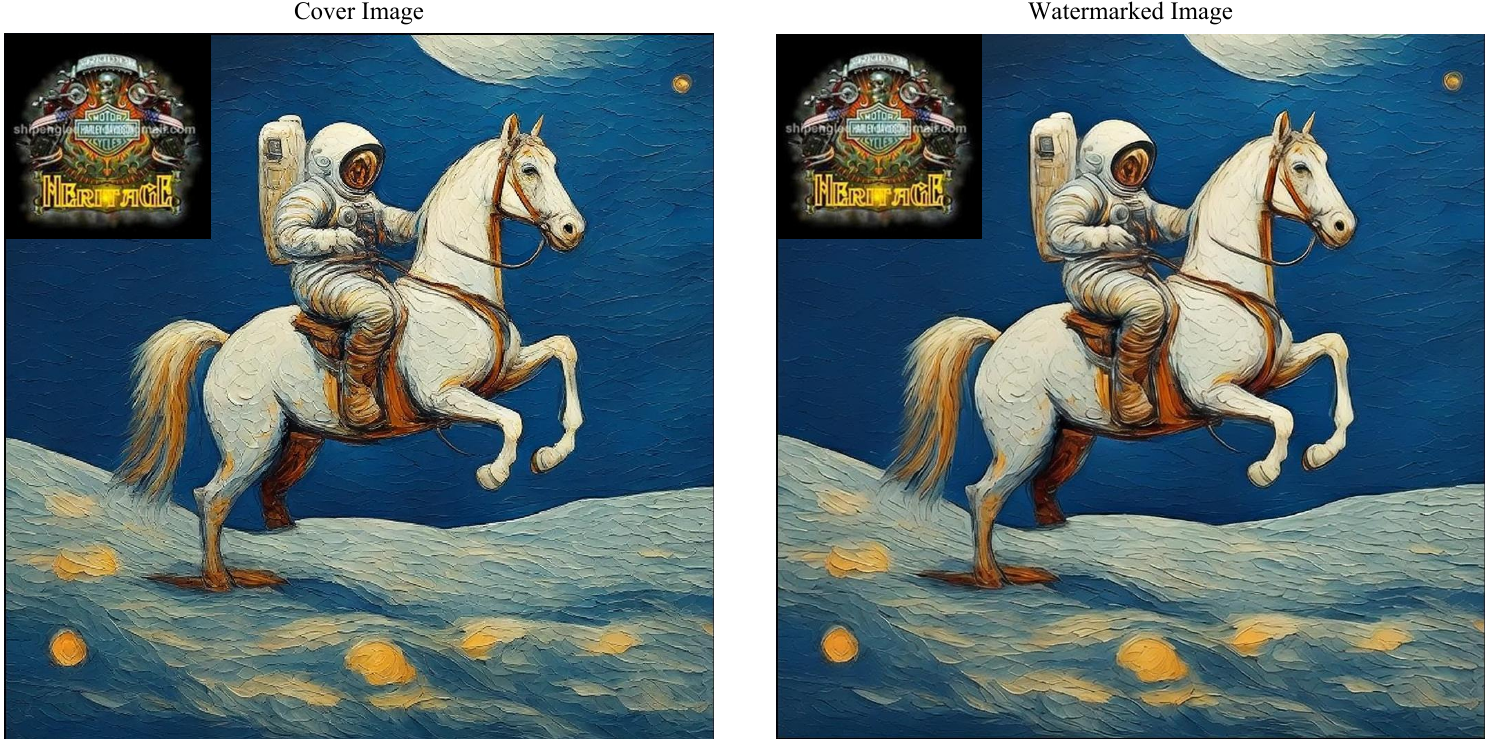}
    \vspace{-3mm}
    \caption{Prompt: An astronaut riding a horse on the moon, oil painting by Van Gogh.
    }
    \label{fig: 1}
    \vspace{-4mm}
\end{figure*}

\begin{figure*}[ht]
    \centering
    \includegraphics[width=0.99\linewidth]{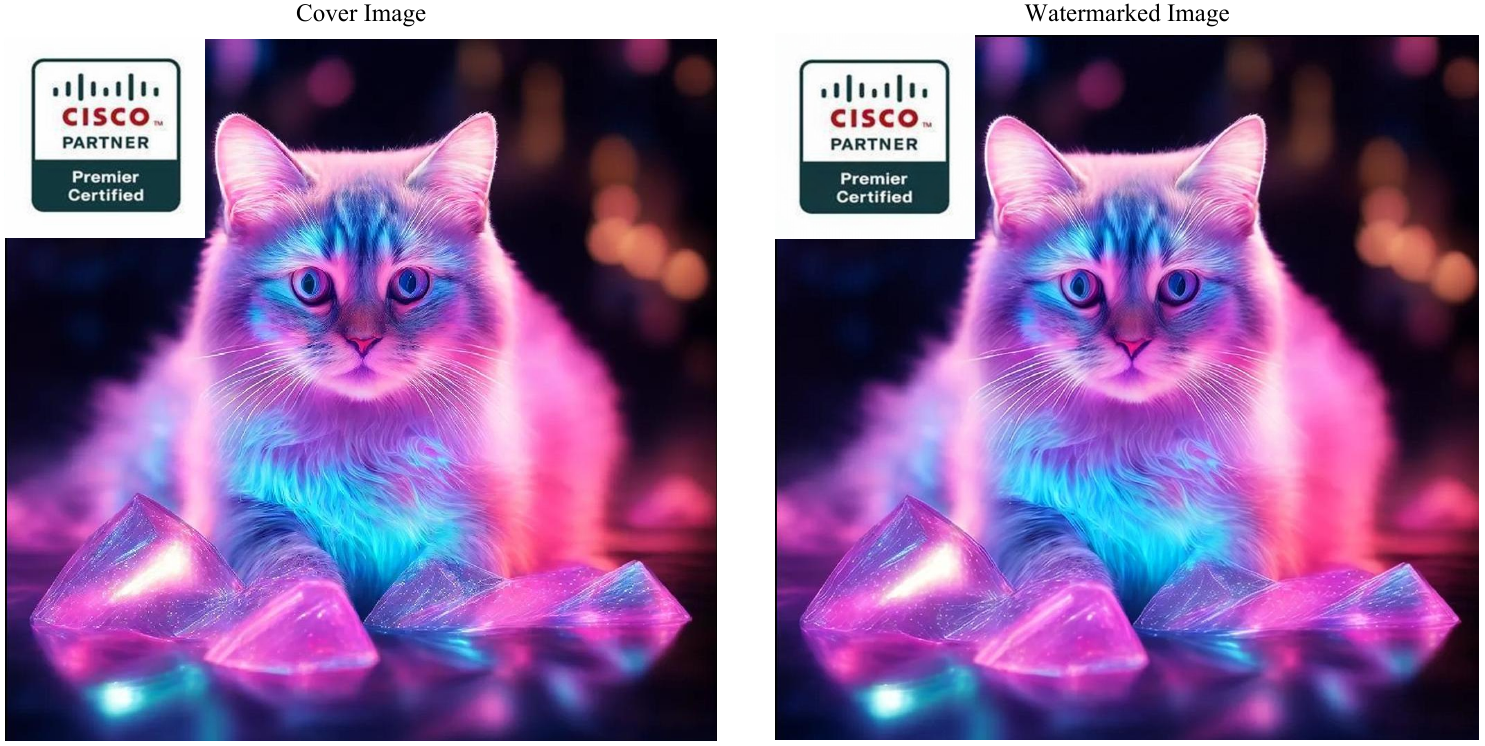}
    \vspace{-3mm}
    \caption{Prompt: A neon holography crystal cat.
    }
    \label{fig: 2}
    \vspace{-4mm}
\end{figure*}
\section{Experimental Results on Autoregressive Text-to-Image Models}

We evaluate \name{} on autoregressive (AR) text-to-image models, demonstrating its effectiveness in embedding watermarks directly during the image generation process. Unlike post-hoc watermarking methods that modify images after generation, our approach seamlessly integrates watermarks at the latent feature level, ensuring both high visual fidelity and robust traceability.

Experimental results show that \name{} consistently achieves superior image quality and watermarking fidelity. Notably, it achieves superior performance even at a high resolution of $1024\times1024$, preserving both image quality and watermarking fidelity. Results are presented in \cref{fig: 1} and \cref{fig: 2}.